\documentclass[twocolumn,aps,prd,nofootinbib,superscriptaddress,showkeys]{revtex4-1}
\usepackage[colorlinks=false,linkcolor=black,citecolor=black]{hyperref}
\usepackage{graphicx}
\usepackage{subfigure}
\usepackage{amstext,amsmath,amsfonts,amssymb,amsthm}
\usepackage[format=plain,justification=centerlast]{caption}
\usepackage{hyperref} 
\usepackage{color}

\def \nn  {\nonumber}

\def \be {\begin{equation}}
\def \ee {\end{equation}}
\def \ee {\end{equation}}

\def \L  {\left[}
\def \R  {\right]}
\def \pa  {\partial}

\def \f   {\frac}

\def \i   {\int}

\def \nb {\nabla}

\begin{document} 

\title{Schwinger Effect in Compact Space}
\date{\today}
\author{Prasant Samantray}
\email{prasant.samantray@hyderabad.bits-pilani.ac.in}
\affiliation{Department of Physics, BITS-Pilani Hyderabad Campus, Jawahar Nagar, Shamirpet Mandal, Secunderabad 500078 India}

\author{Suprit Singh}
\email{suprit@iitd.ac.in}
\affiliation{Department of Physics, Indian Institute of Technology Delhi, New Delhi India}

\begin{abstract}
We consider a theory of scalar QED on a spatially compact $1+1$-dimensional spacetime. By considering a constant electric field pointing down the compact dimension, we compute the quantum effective action by integrating out the scalar degrees of freedom in the Euclidean sector. Working in the saddle-point approximation we uncover two novel branches/physical regimes upon analytically continuing back to real time and discover a new result, hitherto unreported in previous literature. Implications of our results are discussed. 
\end{abstract}

\maketitle

\section{Background}
\label{sec:intro}

It is a well known result of quantum field theory in external backgrounds that strong fields can lead to particle creation from vacuum. The first such prediction was in quantum electrodynamics, of pair creation by strong electric fields, computed by Schwinger in 1951 \cite{Heisenberg:1935qt,Schwinger:1951a}. In almost 70 years that have passed, this effect still awaits confirmation in laboratories amidst ongoing efforts. This is in part due to the very nature of the phenomenon that even the leading order term is exponentially suppressed in flat spacetime. For an excellent general overview the reader is encouraged to look up \cite{nucleation,Dunne:2004nc,Dunne:2010,Affleck:1982,Affleck:1982a,Srinivasan:1999,Padmanabhan} and references therein. Additionally, with the hope of enhancing the effect and other implications, the mechanism has been probed at finite temperatures\cite{Medina,Kim1,Kim2,Gies1,Gies2}, and also in curved spacetimes such as de Sitter and Anti de Sitter backgrounds \cite{dsads}. 

We consider Schwinger effect in $1+1$ dimensional flat spacetime with compact spatial dimension. This idea has been probed before \cite{AdamB,Qiu} but not with the full rigour of computing the effective action and ascertaining particle creation from it. There also appears to be a lot of confusion in the literature specifically in the context of compact spatial dimensions or in the presence of a thermal bath when considering particle production via the Schwinger process. With an aim to demystify this issue, we work this out for scalar quantum electrodynamics and in doing so obtain a new result which has been absent in the previous analysis.
Finally, to answer the question as to why consider the effect in a compact dimension, the extra (large) dimensions are often invoked in, for instance, Braneworld models (see references \cite{Wu1,Wu2,Lu} for string inspired models) and it is necessary we devise tests for probing these extra dimensions. With this motivation, we consider in this article a toy $1+1$ dimensional model which is analytically tractable and will be the basis for future explorations.

\section{Scalar QED in $1+1$ flat Spacetime with compact spatial dimension}
\label{sec:calc}

\noindent Consider a $(1+1)$ dimensional Minkowski space $ds^2 = -dt^2 + dx_1^2$ with compact spatial dimension under the identification $x_1 \sim x_1 + L$. The action for a complex scalar field coupled to an electromagnetic field reads:
\be
S_\phi =  -\i \phi^* \L  -D_\mu D^\mu + m^2\R \phi ~ dtdx_{1} 
\ee
where $D_\mu = \pa_\mu - iqA_\mu$. We shall work in Euclidean time, $x_0 = i t$ such that $ds_E^2 = \delta_{ab} dx^a dx^b = dx_0^2 + dx_1^2$ and $F_{tx} = \pa_t A_{x_1} - \pa_{x_1} A_t \rightarrow i(\pa_{x_0} A_{x_1} -  \pa_{x_1} A_{x_0}) = iF_{x_0 x_1}$. The Euclidean action (labeled by subscript $E$) is then given by
\begin{align}
S^E = & \i dx_0 dx_1 \left[\phi^*(-\nb^2 + m^2)\phi + 2iq\phi^* {\bf A}\cdot \nb\phi \right. \nonumber\\
 &~~\left.+ i q |\phi|^2 \nb\cdot {\bf A} + q^2|\phi|^2 A^2 \right]
\end{align}
with the definition $\nb \equiv (\pa_{x_0},\pa_{x1})$. We now make the gauge choice $A_{x_0} = 0$; $A_{x_1} = E x_0 + a$ respecting the periodic boundary conditions where $a\in R$ cannot be gauged away to zero by any gauge transformation (which respects periodic boundary conditions). Together these conditions imply the Euclidean version of Lorenz gauge, that is, $\nb\cdot{\bf A} = 0$. With this choice, the action becomes
\begin{align}
S^E =& \i dx_0 dx_1 \left[\phi^*(-\nb^2 + m^2)\phi  \right. \nonumber\\
&\left.+2iq\phi^* (Ex_0+a)\pa_{x_1}\phi +q^2|\phi|^2(Ex_0+a)^2 \right].
\end{align}
In order to compute the effective action, it is a standard procedure to integrate out the scalar field in the path integral which formally yields
\be
S^E_{\rm eff} = \ln \det \hat{D} = {\rm Tr} \ln \hat{D} = \i_0^{\infty} \f{ds}{s}\, {\rm Tr} (e^{-s \hat{D}})
\ee
where the operator $\hat{D}$ can be read off from the action of the scalar field to be
\be
\hat{D} = -\nb^2 + m^2 + 2iq(Ex_0+a)\pa_{x_1} + q^2(Ex_0 +a)^2. 
\ee
Since the differential operator $\hat{D}$ is invariant under translations in the spatial dimension and the fact that the eigenfunctions of $\hat{D}$ need to obey the periodic boundary conditions in coordinate $x_1$, we can factorize the eigenfunctions (with proper normalization) as 
\be
\f{1}{\sqrt{L}} f_n(x_0)\exp\left(-\f{2\pi i\, n\, x_1}{L}\right) ~~~\forall n\in \mathbb{Z}.\nonumber
\ee
This yields
\be
\left[-\f{\pa^2}{\pa x_0^2} + q^2 E^2 (x_0 + \f{2\pi n}{q E L} +\f{a}{E})^2 + m^2\right] f_n(x_0) = \lambda f_n(x_0)
\ee
and defining $y := (x_0+2\pi n /qeL + a/E)$ puts it in a more familiar form 
\be
\left[-\f{\pa^2}{\pa y^2} + q^2 E^2 y^2 \right] g_n (y) = (\lambda - m^2) g_n(y)
\ee
which is that of a quantum harmonic oscillator and since $-\infty < x_0 <\infty$, we have $-\infty<y<\infty$ for any fixed $n\in\mathbb{Z}$. By identifying $\omega = 2 q E$ as the frequency of the oscillator, the solutions to this equation can therefore be written straightaway in terms of the wavefunctions of a quantum harmonic oscillator as
\be
\f{1}{\sqrt{L}} \Psi_{nj}\left(x_0 + \f{2\pi n}{qEL} +\f{a}{E}\right) \exp(-\f{2\pi i n x_1}{L}) \nonumber
\ee
with eigenvalues $\lambda_{j,n} = m^2 + (2j+1)qE$  where $j=0,1,2,\ldots \forall n\in \mathbb{Z}$.  Since $\lambda_{j,n}$ are explicitly independent of $n$, the eigenfunctions are infinitely degenerate. 
The heat kernel for the operator $\hat{D}$ can therefore be written as,
\begin{align}
K(s; x_0',x_1' & |x_0,x_1) = \f{1}{L} \sum_{n=-\infty}^{\infty}\sum_{j=0}^{\infty} e^{-s\lambda_{j,n}} \Psi_{n,j}^*(y')\Psi_{n,j}^*(y) \times\nonumber\\
&\phantom{phantom}\times \exp\left(-\f{2\pi i n}{L} (x_1-x_1')\right) \nonumber\\
=& \sum_{n=-\infty}^{\infty} k_n(s;x_0'|x_0) \exp\left(-\f{2\pi i n}{L} (x_1-x_1')\right)
\end{align}
where $k_n$ is the propagator for the quantum harmonic oscillator for a fixed $n$ in Euclidean time. Since the computation of effective action involves taking a trace, we can set $x_0' = x_0$ and $x_1'=x_1$ in the heat kernel which gives
\be
k_n (s;x_0|x_0)= \frac{1}{L} e^{-m^2 s} \left(\f{qE}{2\pi \sinh(2 qE s)}\right)^{\f{1}{2}} e^{-qE \tanh(qEs) y^2}
\ee
and the complete heart kernel (in the coincidence limit) is given by
\begin{align}
K(s;x_0,x_1|x_0,x_1) =& \f{1}{L} e^{-m^2 s} \left(\f{qE}{2\pi \sinh(2 qE s)}\right)^{\f{1}{2}} \times\nonumber \\
&~~\times\sum_{n=-\infty}^{\infty} e^{-qE \tanh(qEs) y^2}
\end{align}
where $y = (x_0+2\pi n /qeL + a/E)$ as defined earlier. Using the Poisson summation formula, this can be manipulated to give 
\begin{align}
&K(s; x_0,x_1|x_0,x_1) = \f{qE}{4\pi} \f{e^{-m^2 s}}{\sinh(qEs)}\times \nonumber\\
&\times\sum_{n=-\infty}^{\infty} \exp \left(-\f{n^2 L^2 qE \coth(qEs)}{4} -iqn(Ex_0+a)L\right).
\end{align}
With this we can write the Euclidean effective action as 
\begin{align}
S_{\rm eff}^E =& -\int_0^\infty \f{ds}{s} \rm{Tr}(e^{-s \hat{D}}) \nonumber \\
=& -\i dx_0dx_1 \i_0^{\infty} \f{ds}{s} K(s;x_0,x_1|x_0,x_1) \nonumber \\
=& \i dx_0dx_1 L_{\rm eff}^E
\end{align}
from which we can read off the effective Lagrangian
\begin{align}
L_{\rm eff}^E =&  -\sum_{n=-\infty}^{\infty} e^{-iqnL(Ex_0+a)} \i_0^\infty \f{ds}{4\pi s^2} e^{-m^2s} \f{qEs}{\sinh qEs} \times\nonumber\\
&\times\exp \left(-\f{n^2 L^2 qE \coth(qEs)}{4} \right).
\end{align}
Interestingly the pre-factor before the integral is actually $\exp(-i q\oint {\bf A}\cdot {\bf dx})$ and corresponds to geometric phase. This geometric phase from its very form is gauge invariant and is the result of the fact that the Euclidean topology considered here is not simply connected. This non-trivial holonomy results in the pre-factor as above. Similar pre-factor appears even in computation of Casimir energy in compact spacetimes, where the integer $n$ takes specific values leading to the minimization of the effective potential that gives the vacuum energy density of the spacetime (details can be found in ref.~\cite{ford}).This term is usually not reported in the literature. We shall drop this term for the calculations in this article in order to focus on the pair production rate. With a change of variables $\tau = qEs$ and defining parameters $\lambda_E = m^2/qE$, $R_E = m/qE$ and $\sigma_E = L/2R_E$ (where now the subscripts indicate Euclidean time) and splitting the $n=0$ and $n\neq0$ terms, we have
\begin{align}
L_{\rm eff}^{E} &= -\f{qE}{4\pi} \i_0^\infty  \f{d\tau}{\tau\sinh\tau} e^{-\lambda_E \tau} \nn\\
 &~~~-\f{qE}{2\pi}\sum_{n=1}^{\infty}  \i_0^\infty  \f{d\tau}{\tau\sinh\tau}  e^{-\lambda_E(\tau+n^2 \sigma^2_E \coth\tau)}\nn\\
 &=L_{\rm eff}^{E\,(0)}  + L_{\rm eff}^{E\,(n>0)}.
\end{align}
This will be analytically continued back to real time with $t = -ix_0$, $E_{\rm phy} = iE$, $\sigma = i\sigma_E$ and $\lambda = -i\lambda_E$ where the latter two imply $\sigma_E\lambda_E = \sigma\lambda$. Considering the first term (which is also independent of $\sigma_E$) after analytically continuing it back to real time, 
\be
L_{\rm eff}^{(0)}  =  \f{qE_{\rm phy}}{4\pi} \i_0^\infty  \f{d\tau}{\tau\sin\tau} e^{-\lambda\tau} 
\ee 
we see that there exist simple poles at $\tau = k\pi$ with $k=1,2,\ldots$. The pole at $\tau = 0$ does not contribute as it is cancelled by the background shift in the limit of vanishing electric field. Therefore, to evaluate the integral we can express $\tau = k\pi + \epsilon \exp(i\theta)$ in the complex-$\tau$ plane. The pole at $\tau=0$ does not contribute to the imaginary part as it is cancelled by the background in the limit the electric field goes to zero. The contour individually circles each of the poles that lie on the real axis which contribute to give
\be
L_{\rm eff}^{(0)} = \f{iqE_{\rm phy}}{4\pi } \sum_{k=1}^{\infty} \frac{(-1)^{k+1}}{k} e^{-\lambda k \pi}.
\ee
This is the standard expression for the particle production rate in flat $1+1$-dimensional spacetime without any compactification. As for the second term, we can use the saddle point approximation (only in the limit where $\lambda_E \gg 1$) to evaluate this Laplace type integral which is of the form
\be
I = \i_a^b g(\tau) e^{-\lambda_E f(\tau)} \simeq  \sqrt{\f{2\pi}{\lambda_E f^{''}(\tau_0)}} g(\tau_0) e^{-\lambda_E f(\tau_0)}
\ee
with $g(\tau) = 1/\tau \sinh \tau$ and $f(\tau) = \tau + n^2 \sigma_E^2 \coth\tau$. Noting that,
\be
f^{'} (\tau) = 1- \f{n^2\sigma_E^2}{\sinh^2(\tau)}; \quad f^{''}(\tau) = \f{2 n^2\sigma_E^2 \cosh\tau}{\sinh^3(\tau)}\nn
\ee
the saddle points from the zeroes of the derivative, $f^{'} (\tau_0) = 0$  are given $\sinh\tau_0 = n\sigma_E$ and hence $e^{\tau_0} = n\sigma_E + \sqrt{1+n^2\sigma_E^2}$ for $n=1, 2, 3, \ldots$. Equivalently we could have solved for the saddle points explicitly as 
$\tau_0 = \log(n\sigma_E + \sqrt{1+n^2\sigma_E^2})$. Therefore, the effective Lagrangian, by summing over the contributions from all the saddle points is 
\begin{align}
L_{\rm eff}^{E\, (n>0)}& \simeq -\f{qE}{2\pi} \sum_{n=1}^{\infty} \sqrt{\f{2\pi}{\lambda_E f^{''}(\tau_0)}} g(\tau_0) e^{-\lambda_E f(\tau_0)}\nn\\
&= -\f{qE}{\sqrt{4\pi\lambda_E}} \sum_{n=1}^{\infty} \f{e^{-\lambda_E[n\sigma_E\sqrt{1+n^2\sigma_E^2} +\sinh^{-1}n\sigma_E]}}{\sqrt{n\sigma_E}\sinh^{-1}n\sigma_E (1+n^2\sigma_E^2)^{1/4}}
\end{align}
which will be our master formula. Modulo the choice of gauge (in accordance with the non-trivial holonomy), it is evident that this formula is mathematically unambiguous and therefore unique in $1+1$-dimensional compact space. We now have two cases of physical interest to consider.

\subsection{Case: $\sigma>1$ or $L>2R_0$} 

The correct analytic continuations are 
\be
\sqrt{1+n^2\sigma_E^2} \rightarrow -i \sqrt{n^2\sigma^2 -1}; \nn
\ee
\be
\sinh^{-1}(in\sigma) = i \pi/2 - \cosh^{-1} n\sigma \nn
\ee
The choice of these analytic continuations are due to the following considerations. Consider a complex multivalued function $f(z) = \log(z + \sqrt{1 + z^2})$. In order to make this function analytic, we choose a branch cut starting at $z=-i$ and running all the way upwards along the imaginary axis to $z = i\infty$ in the complex z-plane. This choice of the cut ensures that $f(z)\in \mathbb{R}$ for $z \in \mathbb{R}$. Plugging this into the master formula we get 
\begin{align}
&L_{\rm eff}^{(n>0)} = \f{iqE_{\rm phy}}{\sqrt{4\pi\lambda \sigma}} \times\nn\\
&~~\times\sum_{n=1}^{\infty} \f{e^{-i\lambda[-n\sigma\sqrt{n^2\sigma^2-1} -i\pi/2 + \cosh^{-1}n\sigma]}}{\sqrt{n}(-i\pi/2+\cosh^{-1}n\sigma) e^{-i\pi/4}(n^2\sigma^2-1)^{1/4}}
\end{align}
Setting $n\sigma = \cosh(\alpha_n)$, since $\sigma>1$ and $n =1,2,\ldots$ we have 
\be
L_{\rm eff}^{(n>0)} = \f{iqE_{\rm phy}}{\sqrt{4\pi\lambda \sigma}} e^{i\pi/4} \sum_{n=1}^{\infty} \f{e^{-i\lambda \alpha_n + i(\lambda/2) \sinh (2\alpha_n) - \lambda \pi/2}}{\sqrt{n\sinh\alpha_n} (-i\pi/2+\alpha_n)}.
\ee
This can further be split (noting that $\lambda\sigma = mL/2$) as 
\begin{align}
L_{\rm eff}^{(n>0)}&= \f{iqE_{\rm phy}}{\sqrt{2\pi mL}} e^{-\lambda\pi/2} \sum_{n=1}^{\infty} \f{[(\pi/2) \sin\gamma_n + \alpha_n \cos\gamma_n]}{\sqrt{n\sinh\alpha_n} (\pi^2/4+\alpha_n^2)} \nn\\
&+ {\rm Re}(L_{\rm eff}).
\end{align}
where $\gamma_n = \lambda\alpha_n - (\lambda/2)\sinh(2\alpha_n) -\pi/4 $. 
The complete effective Lagrangian for $\sigma>1$ is,
\begin{align}
L_{\rm eff} &= \f{iqE_{\rm phy}}{4\pi } \sum_{n=1}^{\infty} \frac{(-1)^{n+1}}{n}e^{-\lambda n\pi} + \f{iqE_{\rm phy}}{\sqrt{2\pi mL}} e^{-\lambda\pi/2} \times \nn\\
&\times\sum_{n=1}^{\infty} \f{[(\pi/2) \sin\gamma_n + \alpha_n \cos\gamma_n]}{\sqrt{n\sinh\alpha_n} (\pi^2/4+\alpha_n^2)} + {\rm Re}(L_{\rm eff})
\end{align}
Further for $\sigma\gg1$, only the $n=1$ order term contributes maximally so that
\be
L_{\rm eff} \approx \f{iqE_{\rm phy}}{4\pi }  e^{-\lambda \pi} + \f{iqE_{\rm phy}}{\sqrt{2\pi mL}} e^{-\lambda\pi/2}  \f{1}{\sqrt{\sigma}\ln\sigma} + {\rm Re}(L_{\rm eff}).
\ee
Since the vacuum to vacuum expectation value is,
\be
\langle \rm{out} | \rm{in}\rangle = \exp[iS_{\rm eff}] = \exp[iLTL_{\rm eff}] \nn
\ee
where $L$ is the volume of the compact space, and $T$ is the time interval under consideration. The pair production probability which is $\mathbb{P} = 1- |\langle \rm{out} | \rm{in}\rangle|^2 $ picks up only the imaginary part of the effective Lagrangian giving
\be
\mathbb{P}  \approx LT\left[\f{qE_{\rm phy}}{2\pi }e^{-\lambda \pi} +\f{2qE_{\rm phy}}{\sqrt{2\pi mL\sigma}\ln\sigma} e^{-\lambda \pi/2}\right]. \label{central result}
\ee
This is an entirely new result valid for $\sigma\gg1$ (along with $\lambda\sigma \gg 1$) that is in the case of a large compact dimension, the $\sigma$-dependent term appears to enhance the standard Schwinger pair creation rate. Further, as a consistency check, in the limit $\sigma \rightarrow \infty$ it reduces to the standard expression for Schwinger effect in $1+1$ Minkowski spacetime with flat topology. 
 
\subsection{Case: $\sigma<1$ or $L<2R_0$} 

In this case the correct analytic continuations are
\begin{align}
~~~~\sqrt{1+n^2\sigma_E^2} &\rightarrow \sqrt{1-n^2\sigma^2}; \nn\\
-\sinh(in\sigma) = \sinh(n\sigma_E) &\rightarrow -i[\sin^{-1} n\sigma +2k\pi] \nn
\end{align}
with $k=0,1,2,\ldots$ for $n \sigma<1$ with $n = 1, 2, 3, \ldots, [1/\sigma]$, where $[x]$ is the greatest integer function. Plugging this into the master formula we get 
\begin{align}
&L_{\rm eff}^{(n>0)} = \f{qE_{\rm phy}}{\sqrt{4\pi\lambda}}\times \nn\\
&\times \sum_{k=1}^{\infty} \sum_{n=1}^{n_{\rm max}} \f{e^{-\lambda[n\sigma\sqrt{1+n^2\sigma^2} +\sin^{-1}n\sigma + 2\pi k]}}{\sqrt{n\sigma}[\sin^{-1}(n\sigma) + 2\pi k] (1+n^2\sigma^2)^{1/4}}.
\end{align}
The total effective Lagrangian is, therefore,
\begin{align}
&L_{\rm eff} =  i\f{qE_{\rm phy}}{4\pi} \sum_{n=1}^{\infty}\frac{(-1)^{n+1}}{n} e^{-\lambda n\pi}\nn\\
&~~+ \f{qE}{\sqrt{4\pi\lambda}} \sum_{k=1}^{\infty} \sum_{n=1}^{n_{\rm max}} \f{e^{-\lambda[n\sigma\sqrt{1+n^2\sigma^2} +\sin^{-1}n\sigma + 2\pi k]}}{\sqrt{n\sigma}[\sin^{-1}(n\sigma) + 2\pi k] (1+n^2\sigma^2)^{1/4}}.
\end{align}
where $n_{\rm max}=[1/\sigma]$. It is noteworthy that only the first term contributes to the imaginary part of the effective action giving particle production that is independent of $\sigma$ and hence the compactification scale $L$ in this case and the Schwinger mechanism is identical to that in normal flat spacetime. The second term is real and therefore goes into correcting the Maxwell's action in $1+1$-dimensional spacetime. Also, the $n=1$ lowest order term coincides with results discussed in \cite{AdamB,Kleban} in the context of flux discharge and decay of the electric field.

\section{Discussion}
Schwinger effect, that is, particle production due to vacuum decay induced by external fields is still one of the outstanding radical predictions of quantum electrodynamics and quantum field theory in external backgrounds in general. In fact, the mechanism in its various appearances is also a basis for various theoretical strides in gravitation and cosmology, and even theoretical condensed matter physics. The inverse exponential dependence on external Electric field suppresses the mechanism requiring strong fields. A way forward is to seek arenas where enhancement of the effect can occur with some augmentation over the usual Poincare invariant flat spacetime. We consider the effect in scalar quantum electrodynamics in $1+1$ dimensional flat spacetime with compact spatial dimension. The motivation for this is twofold, one as a check for indications of enhancement and second being a step towards devising tests for large extra dimensions invoked in string inspired models.\\
\\
We find through rigorous computation of the effective action results previously unstated in the literature. The calculations were done in the Euclidean sector with zero ambiguities. Remarkably, the Euclidean sector contains information about both regimes of interest viz. (a) $\sigma<1 \Rightarrow L<2m/qE$ and (b) $\sigma>1 \Rightarrow L>2m/qE$ when rotated back to real time. There is a standard term (independent of compactification scale or "$\sigma$") that always contributes to particle production in both cases. For case (a) the compactification has no effect on the rate of particle production. Certain previous studies have reported that there is no particle production in this regime, since in the usual instanton approach the energy density of the electric field does not drop to the requisite amount to create an on-shell real particle pair. In this picture a virtual pair is created at a spatial point and then separate in imaginary time until they meet up at the antipodal point and annihilate. However, this picture of energy conservation is heuristic and possibly incomplete in the context of small compact spatial dimensions with non-trivial holonomy. On the other hand, for case (b) we predict a new rate obtained via analytic continuation from the Euclidean sector. This can have implication for the early universe cosmology and braneworld models. We note that there is a discontinuity at $L=2m/qE$ upon analytic continuation back to real time. This can attributed to the saddle point approximation and possibly the situation will be remedied if a complete quantum or numerical analysis is performed. It is important to note that a central tenet of our analysis is that we assume $\sigma_E\lambda_E = \sigma\lambda \gg 1$ for both the regimes. This is turn translates to the simple condition that $mL \gg 1$, which basically implies that the compactification scale is much larger than the compton wavelength of the particle, thereby allowing a semi-classical analysis to hold. Regimes where $mL \ll 1$ would have to be treated by considering the full effects of backreaction on the gauge field sector as well. \\
\\
Additionally, our results can be generalized in a straightforward manner even to the Schwinger mechanism at finite temperature but without any spatial compactification. This is possible since from the perspective of the Euclidean sector it is just relabelling of the coordinate axes and we would have to identify the Euclidean time coordinate $x_0 \sim x_0 + \beta$ instead. We can then read off the effective lagrangian by switching $L \rightarrow \beta$. We rediscover the temperature regimes $T>T_c$ and $T<T_c$ reported previously in the literature \cite{AdamB,Medina} where $T_c = 1/R_0=qE/2m$. The interesting thing to note is that for $T>T_c$, there appears to be no effect on the Schwinger process and the $n=0$ term in our master formula remains the relevant rate. This is completely in line with results of \cite{Gies1,Gies2}, However for the regime $T<T_c$ our calculations show that thermality does contribute to the one-loop effective action previously unreported in the literature. This sharp behavior at $\sigma=1$ (for spatially compact case) and for $T=T_c$ (in the thermal case) is likely indicative of either (1) the need to go beyond the one-loop order and beyond the saddle point approximation while evaluating the master formula, or  (2) an intricate structure of the effective action owing to the choice of gauge respecting the periodic boundary conditions (thereby leading to the non-trivial holonomy) and working entirely in the Euclidean formalism. There appears not to be any general consensus on this yet in the present literature. For more references on this contentious issue please refer to \cite{Cox,Rojas,Elmfors,Ganguly,Hallin,Gould1,Gould2,Gould3,Draper,Greger, ArunT}. \\
\\
Our analysis has implications for models of cosmology which have large extra dimensions as a feature. Even though our analysis was done strictly in 1+1 dimensions, we can nevertheless draw some qualitative inferences. From Eq(\ref{central result}), we can see that there is an exponential enhancement for the rate of particle production as the compactification scale approches $\sigma \rightarrow 1$ from the above. This indicates that as the large extra dimensions got compactified in the inflationary era via some dynamical mechanism, the universe was populated with charged particles at an exponentially faster rate compared to the usual Schwinger result without any extra dimensions. This would in turn have resulted in a faster shorting of the electric field at extra-galactic scales.  With production of charged particles enhanced via both the inflationary expansion, as well as through the mechanism stated in Eq(\ref{central result}), this would have rapidly increased plasma conductivity in early inflationary era, thereby influencing the generation and evolution of primordial electromagnetic fields. For astrophysical implications the reader is urged to look at \cite{Kandu} and references therein. Additionally as remarked earlier, since we worked entirely in the Euclidean sector, our result of Eq(\ref{central result}) is applicable for finite temperature systems as well with the identification that $L = 1/T$. Therefore, a more earthly application of our result can be in the context of bilayer graphene \cite{AdamB} with a mass-gap of few hundred Mev ($\sim 200$ Mev), a total voltage drop across a sample size (8-10 microns) of the order around 500-600 Mev, and at a temperature $T \sim 200K$ so that we have $T<T_c$. With modern technology one should in principle be able to test the validity of our result by directly measuring the rate of particle production.\\  
\\
It is important to note that we did not consider backreaction in our analysis and that issue is still outstanding. For one it is not clear how the imaginary term of the effective action and hence particle production can be accounted for in modifying the classical equations for the background without any artificial manipulations. A self-consistent approach to handle backreaction is still pending possibly through semi-analytical and/or numerical techniques. This will also seek to answer the question, what is the fate of QED vacuum in compact space by incorporating backreaction as well as inhomogenous fields on compact spacetime including a study in higher than two dimensions considered in this article.

\section{Acknowledgments}

Research of P.S is partially supported by CSIR grant 03(1350)/16/EMR-II Govt. of India, and also by the OPERA fellowship from BITS-Pilani, Hyderabad Campus. Research of S.S. is partially supported by Young Faculty Incentive Fellowship from IIT Delhi, India.

\end{document}